%% file: IRF_four_blocks_v3_fix.tex
\input phyzzx
\input mydef
\input tikz.tex
\def\center#1\endcenter{\centerline{#1}}
\def\threeline#1#2#3{\displaylines{\qquad#1\hfill\llap{(\adveq\myeq)}\cr\hfill#2\hfill \cr
\hfill#3\qquad\cr}}
\def\fourline#1#2#3#4{\displaylines{#1\hfill\llap{(\adveq\myeq)}\cr#2\hfill \cr\hfill#3\hfill\cr
\hfill#4\qquad\cr}}
\def\fiveline#1#2#3#4#5{\displaylines{#1\hfill\llap{(\adveq\myeq)}\cr#2\hfill \cr\hfill#3\hfill\cr\hfill#4\cr
\hfill#5\qquad\cr}}

\date{January,  2019}
\date{January, 2019}
\titlepage
\title{On the Algebraic Approach to Solvable Lattice Models}
\author{Vladimir  Belavin$^{a,b,c}$ and Doron Gepner$^a$}
\vskip20pt
\line{\it\hfill  $^a$\  Department of Particle Physics and Astrophysics, Weizmann Institute, \hfill}
\line{\it\hfill Rehovot 76100,  Israel\hfill} 
\line{\it \hfill $^b$ \  I.E. Tamm  Department of Theoretical Physics, P.N. Lebedev Physical
\hfill }\line{\hfill\it  Institute, Leninsky av. 53, 11991 Moscow, Russia\hfill}
\line{\it \hfill $^c$ \   Department of Quantum Physics, Institute for Information Transmission\hfill}
\line{\hfill\it  Problems, Bolshoy Karetny per. 19,  127994 Moscow, Russia\hfill}

\abstract
We treat here interaction round the face (IRF) solvable lattice models. We study the algebraic 
structures underlining such models. For the three block case, we show that the Yang Baxter
equation is obeyed, if and only if, the Birman--Murakami--Wenzl (BMW) algebra is obeyed. We prove this
by an algebraic expansion of the Yang Baxter equation (YBE). For four blocks IRF models, we show
that the BMW algebra is also obeyed, apart from the skein relation, which is different.
This indicates that the BMW algebra is a sub--algebra for all models with three or more blocks.
We find additional relations for the four block algebra using the expansion of the YBE.
The four blocks result, that is the
BMW algebra and the four blocks skein relation, is enough to define new knot
invariant, which depends on three arbitrary parameters, important in knot theory. 
\endpage 

\mysec{Introduction}
Our interest in this paper is solvable interaction round the face (IRF) lattice models in two dimensions.
The solvable lattice models are important playground to study statistical mechanics systems
and their phase structure. For a review see \REF\Baxter{R.J. Baxter, ``Exactly solved models in statistical mechanics", Academic Press, London, England (1982).}\REF\Wadati{M. Wadati, T. Deguchi and Y. Akutsu, Phys. Rep. (180) (4\&5) (1989).}\r{\Baxter,\Wadati}.

The IRF lattice models are strongly connected with two dimensional conformal field theory.
First, the models have a second order phase transition points which are described by some
conformal field theory. Second, the construction of the models themselves is achieved through
the data of some conformal field theories \REF\Found{D. Gepner, ``Foundation of Rational quantum
field theory", arXiv: hep-th/9211100v2 (1992).}\r\Found.
Apart from describing second order phase transitions, the conformal field theories are important
in string theory compactifications, where they describe the world sheet dynamics, see, e.g.,
\REF\GepWit{D. Gepner  and E. Witten, Nucl. Phys. B278 (1986) 493.}\r\GepWit.

Our purpose in this paper is to describe the algebraic structure of solvable IRF lattice models.
We have already started this investigation in a previous paper 
\REF\BelGep{V. Belavin and D. Gepner, ``Three blocks solvable lattice models and Birman--Murakami--Wenzl algebra", arXiv:1807.05603 (2018).}\r\BelGep, where it was argued that any three 
blocks IRF model obeys the Birman--Murakami--Wenzl (BMW) algebra
\REF\BirWen{J.S. Birman and H. Wenzl, Trans. Am. Math. Soc. 313 (1) (1989) 313.}
\REF\Mur{J. Murakami, Osaka J. Math. 24 (4) (1987) 745.}\r{\BirWen,\Mur}.
Here, we wish
to prove this result in detail along with proving that the Yang Baxter equation (YBE) is obeyed
if the BMW algebra holds. 

We then study four block theories and we show that they too obey the BMW algebra, with a 
different skein relation. We exemplify this result by calculating numerically the algebra from
the $3\times 3$ fused $SU(2)$ model \REF\Jimbo{E. Date, M. Jimbo, T. Miwa and M. Okado, Lett.
in Math. Phys. 12 (1986) 209}\r\Jimbo.

This result is enough to generate all the relations of the four block algebra. It is also
sufficient to define a new knot invariant, using the algebraic relations to `reduce' words
in the braid group enveloping algebra. Thus, this algebraic structure is important in knot
theory. For a review on the application of IRF models to knot theory, see \r\Wadati\
and also \REF\GepKnot{D. Gepner, ``On solvable lattice models and knot invariants",
hep-th/9305182 (1993).}\r\GepKnot.

The connection between solvable lattice models and BMW algebra was discussed previously
 \REF\Gallas{W. Galleas and M.J. Martin, arXiv: nlin/04060003 (2004).}
\REF\Vernier{E. Vernier, J.L. Jacobsen and H. Saleur, arXiv: 1404.4497 (2014).}
in connection to the BCD models ref. \r\Wadati, and refs. therein, for certain  superalgebras
\r\Gallas\ and in connection to the Izergin--Korepin model \r\Vernier. We agree with these works and
our results generalize them. 

We hope that our results will further the understanding of solvable lattice models. An
important question is to figure out the algebra underlying the general $n$ block 
lattice model, with arbitrary $n$. We conjecture that the BMW algebra (without the
skein relation) is a sub--algebra for any number of blocks greater or equal three, $n\geq 	3$.

\mysec{Interaction Round the Face lattice models}

We define an Interaction Round the Face (IRF) lattice model from the braiding matrix of
a rational conformal field theory model. The Boltzmann weights obey the limit
$$\lim_{u\rarrow i\infty} g(u)\, \omega\left(\matrix{a & b\cr c&d \cr}\bigg| u\right)=C_{c,d}\left[ \matrix{h &v \cr a & b\cr}\right],\e$$
where $\omega$ is the Boltzmann weight and
$C$ is the Braiding matrix of the conformal field theory 
\REF\MS{G. Moore and N. Seiberg, Phys. Lett. B 212 (1988) 451.}\r\MS, i.e., the braiding of the four point conformal blocks (see Fig. 1), and $g(u)$ is some irrelevant function, used to make the 
limit finite. The primary fields $h$ and $v$ are some fixed primary fields used in the definition of the IRF model and
$a,b,c,d$ are any of the primary fields of the conformal field theory, $\cal O$. The variables 
$z_1$, $z_2$, $z_3$ and $z_4$ are the coordinates of the four point conformal block in the complex plane.
We denote this IRF model as IRF$({\cal O},h,v)$.
\vskip 1cm
\center
\tikzpicture[scale=1]
\draw [line width=1pt] (31.5,0) -- (35.8,0);
\draw [line width=1pt] (32.8,0) -- (32.8,1.8);
\draw [line width=1pt] (34.5,0) -- (34.5,1.8);
\draw [line width=1pt] (41.5,0) -- (45.8,0);
\draw  [line width=1pt] (42.8,0) to[out=90,in=215] node [sloped,above] {} (43.6,0.84);
\draw  [line width=1pt] (43.75,.95) to[out=45,in=-90] node [sloped,above] {} (44.5,1.8);  
\draw  [line width=1pt] (44.5,0) to[out=90,in=-90] node [sloped,above] {} (42.8,1.8);
\draw (30.9,-0) node {$a, z_1$};
\draw (32.6,2.1) node {$ v, z_2$};
\draw (34.5,2.1) node {$ h, z_3$};
\draw (36.3,0) node  {$b, z_4$};
\draw (33.5,-0.4) node {$c$};
\draw (38.8,1.0) node {$=\,\sum_{d} C_{c,d}\left[ \matrix{h &v \cr a & b\cr}\right]\times\,$};
\draw (40.9,-0) node {$a, z_1$};
\draw (42.6,2.1) node {$ h, z_3$};
\draw (44.5,2.1) node {$ v, z_2$};
\draw (46.3,0) node  {$b, z_4$};
\draw (43.5,-0.4) node {$d$};
\endtikzpicture
\endcenter
\vskip 5mm
\center
\Fig 1. { Braiding matrix.}
\endcenter

We find it convenient to define the face transfer matrix as the operator $X_i$,
$$\left< a_1,a_2,\ldots,a_n| X_i(u)|a_1^\prime,a_2^\prime,\ldots,a_n^\prime\right>=\left[\prod_{j\neq i}
\delta_{a_j,a_j^\prime}\right] \omega\left( \matrix{a_{i-1} & a_i \cr a_i^\prime & a_{i+1}} \bigg| u\right).\e$$

Our purpose is to introduce a solution of the Yang Baxter equation (YBE),
$$X_i(u) X_{i+1}(u+v) X_i(v)=X_{i+1}(v) X_i(u+v) X_{i+1}(u),\e$$
which ensures that the transfer matrices for different spectral parameters, $u$, commute. 

The fusion rules of the primary fields $h$ and $v$ enter into the conformal data:
$$[h]\times [v]=\sum_{i=0}^{n-1} \psi_i,\e$$
where $[h]$ and $[v]$ are some primary fields in the CFT $\cal O$. The product is according
to the OPE (fusion rules)  and $n$ is the number of  conformal blocks (for shortness blocks).
The eigenvalues of the braiding matrix are given by,
$$\lambda_i=\epsilon_i e^{i\pi(\Delta_h+\Delta_v-\Delta_i)},\e$$
where $\epsilon_i=\pm1$, according to whether the product is symmetric or antisymmetric.
We shall assume that $h=v$ and $h$ is real, $\psi_0=1$, the unit field, and that $\epsilon_i=(-1)^i$.

Denote by $X_i$ the limit of the Brading matrix as $u\rarrow i\infty$ (up to a factor). We then see that
$X_i$ obeys an $n$th order polynomial equation,
$$\prod_{p=0}^{n-1} (X_i-\lambda_p)=0.\e$$
We define the projector operators by,
$$P_i^a=\prod_{p\neq a} \left[{X_i-\lambda_p\over \lambda_a-\lambda_p}\right].\e$$
The projection operators obey the relations,
$$\sum_{a=0}^{n-1} P_i^a=1,\qquad P_i^a P_i^b=\delta_{a,b} P_i^b,\qquad \sum_{a=0}^{n-1} \lambda_a 
P_i^a=X_i.\e$$

In ref. \r\Found, a conjecture for the trigonometric solution of the
YBE was introduced. To describe it, we define parameters
$$\zeta_i=\pi(\Delta_{i+1}-\Delta_i)/2,\e$$
where $\Delta_i$ is the dimension of $\psi_i$.
The trigonometric solution of the Yang Baxter equation is then,
$$X_i(u)=\sum_{a=0}^{n-1} f_a(u) P_i^a,\e$$
where the functions $f_a(u)$ are 
$$f_a(u)=\left[ \prod_{r=1}^a \sin(\zeta_{r-1}-u)\right] \left[\prod_{r=a+1}^{n-1} \sin(\zeta_{r-1}+u)\right]
\bigg/ \left[ \prod_{r=1}^{n-1} 
\sin(\zeta_{r-1})\right].\e$$

For our purposes the following relations will be relevant. 
First, the Boltzmann weights obey crossing symmetry:
$$\omega\left( \matrix{a&b\cr c& d\cr}\bigg| \lambda-u\right)=\left[{G_b G_c\over G_a G_d}\right]^{1/2}
\omega\left( \matrix{c & a \cr d & b\cr}\bigg| u\right),\e$$
where $G_a$ is the crossing multiplier and $\lambda=\zeta_0$ is the crossing parameter.

Another relation is the inversion relation for the transfer matrices:
$$X_i(u) X_i(-u)=\rho(u) \rho(-u) 1_i,\e$$
where 
$$\rho(u)=\prod_{r=1}^{n-1} {\sin(\zeta_{r-1}-u)\over \sin(\zeta_{r-1})}.\e$$

\mysec{Birman--Murakami--Wenzl algebra}

Our aim is to connect the solvable IRF model with Birman--Murakami--Wenzl algebra
\r{\BirWen,\Mur}. There are two generators of the algebra, $G_i$ and $E_i$.
The relations are,
$$G_i G_j=G_j G_i {\ \rm  if\ }  |i-j|\geq 2,$$
$$G_i G_{i+1} G_i=G_{i+1} G_i G_{i+1},\qquad E_i E_{i\pm1} E_i=E_i,$$
$$G_i-G_i^{-1} =m(1-E_i),$$
$$G_{i\pm1} G_i E_{i\pm1}=E_i G_{i\pm1} G_i=E_i E_{i\pm1},\qquad G_{i\pm1} E_i G_{i\pm1}=
G_i^{-1} E_{i\pm1} G_i^{-1},$$
$$G_{i\pm1 } E_i E_{i\pm1}=G_i^{-1} E_{i\pm1},\qquad E_{i\pm1} E_i G_{i\pm1}=E_{i\pm1} G_i^{-1},$$
$$G_i E_i=E_i G_i=l^{-1} E_i,\qquad E_i G_{i\pm1} E_i=l E_i.\e$$
These relations imply the additional relations,
$$E_i E_j=E_j E_i {\ \rm if \ } |i-j|\geq 2,\qquad (E_i)^2=b E_i,\e$$
where here $b=(l-l^{-1})/m +1$.
Here $l$ and $m$ are the two parameters of the algebra.

Assume now the three block case, $n=3$. We connect out solvable IRF lattice model by defining
$$\manyeq{G_i&=4\sin(\lambda)\sin(\mu) e^{-i\lambda} X_i,\cr
                    G_i^{-1}&=4 \sin(\lambda)\sin(\mu) e^{i\lambda} X_i^t,\cr} $$
where 
$$X_i^t=\lim_{u\rarrow -i\infty} e^{2 i u} X_i(u)\e$$
and the normalization is fixed so that $G_i G_i^{-1}=1_i$ from the inversion relation, eq. (2.13). The phase is arbitrary,
and is fixed to be compatible with the BMW algebra.
We also define $E_i=X_i(\lambda)$.

We propose the following Baxterization of the BMW algebra,
$$U_i(u)=1-{i \sin (u)\over 2 \sin(\lambda) \sin(\mu)} \left [ e^{-i (u-\lambda)} G_i-e^{i(u-\lambda)} G_i^{-1} \right ],\e$$
where we identify $\lambda=\zeta_0$ and $\mu=\zeta_1$.

For three blocks, the face transfer matrix  $X_i(u)$, eq. (2.10),  assumes the form,
$$\twoline{X_i(u)=\big[ P^0_i \sin(\zeta_0+u) \sin(\zeta_1+u)+P^1_i \sin(\zeta_0-u) \sin(\zeta_1+u)+}{P^2_i 
\sin(\zeta_0-u)\sin(\zeta_1-u)  \big] /\left[ \sin(\zeta_0)\sin(\zeta_1)\right].}$$
We can then see that with this definition, the Baxterized BMW algebra, $U_i(u)$, eq. (3.5),
and the face transfer matrix, are identical:
$$U_i(u)=X_i(u).\e$$
We also identify 
$$E_i=U_i(\lambda)=E_i(\lambda).\e$$

Several relations are evident from the definitions of $X_i(u)$ and $U_i(u)$. First,
from the crossing symmetry,  eq. (2.13), we find that
$$E\pmatrix{a & b\cr c& d\cr}=\left( {G_b G_c\over G_a G_d}\right)^{1/2} \delta_{a,d},\ \  {\rm and\  \  } E_i E_{i\pm1} E_i=E_i.\e$$
From the definition of $E_i$ in terms of projection operators, we find,
$$E_i^2=b E_i,\qquad{\rm where \ \ } b={\sin(2\lambda)\sin(\mu+\lambda)\over
\sin(\lambda)\sin(\mu)}.\e$$
Thus $E_i$ obeys the Temperley--Lieb algebra. It is noteworthy that for any number of 
blocks, the Temperley--Lieb algebra is obeyed  with 
$$b=\prod_{r=0}^{n-2} {\sin(\lambda+\zeta_r)\over \sin(\zeta_r)}.  $$

Another relation that is evident is the braiding relation,
$$\manyeq{G_i G_{i+1} G_i&=G_{i+1} G_i G_{i+1},\cr
                   G_i G_j&=G_j G_i{\ \rm if \ } |i-j|\geq 2.\cr}$$

From the definition of $E_i$ we find the skein relation,
$$G_i-G_i^{-1}=m(1-E_i),\e$$
where we identify
$$m=-2 i\sin(\mu),\e$$
as one of the parameters of the BMW algebra.

Another relation, which is evident from the definition of the face transfer matrix, $X_i(u)$, eq. (3.6),
is
$$G_i E_i=E_i G_i=l^{-1} E_i,\e$$
where the parameter $l$ is given by,
$$l=-e^{i(2\lambda+\mu)},\e$$
which is the second parameter of the BMW algebra. We note, in passing, that this relation, 
eq. (3.14) is obeyed by any number of blocks, greater than two,  with some value of $l$.

One can easily calculate 
$$b=(l-l^{-1})/m +1,\e$$
which is the relation required by the BMW algebra.

Once establishing these evident relations, which form part of the BMW algebra, we wish to prove
that the face transfer matrix, $X_i(u)$, obeys the Yang Baxter equation if and only if
$G_i$ and $E_i$ obey the relations of the BMW algebra.

For this purpose, we convert the Yang--Baxter equation to a set of algebraic equations obeyed by
$G_i$ and $E_i$. We do this by inserting $X_i(u)$, eq. (3.6), into the YBE, eq. (2.3). We then 
expand the YBE in terms of $G_i$, $E_i$ and $G_i^{-1}$, in powers of $e^{iu}$ and $e^{iv}$.
We get $19$ equations and solve them in terms of the `basis' elements which is
$B_i B_{i+1} B_i$, where $B_i$ is either $G_i$, $E_i$ or $G^{-1}_i$. We get from this
$19$ equations which are listed in appendix (A).

We wish to show that these equations hold, if and only if, the BMW algebra holds. There are 
$12$ equations which contain a single term only. For example, eq. (A.10) is just the
braiding relation, eq. (3.11), which we know that it holds. Similarly, eqs. (A.1--A.4,A.9), are 
the same braiding relation, where we multiply by $G_i$ or $G_{i+1}$ from the left and right. Thus,
these equations are all equivalent to the braiding relation, eq. (3.11).

The rest of the one--term equations are all equivalent to one equation. In particular, eqs. (A.11,A.19)
are directly seen to be the BMW relation,
$$E_i G_{i\pm 1} G_i=G_{i\pm 1} G_i E_{i\pm 1}.\e$$
The rest of the relations, eqs. (A.4,A.6,A.11,A.12,A.17) are then seen to be equivalent to this 
relation, eq. (3.17), by multiplying the l.h.s. and the r.h.s. by the algebra elements
$G_i$ or $G_{i+1}$.

We get now to the $7$ `composite' relations  (that is  having more than one term). These 
vary in complexity. First consider the relation (A.5). Using the skein relation, eq. (3.12),
we substitute in this relation $G_j^{-1}\rarrow -m+m E_j+G_j$, where $j=i$ or $i+1$.
Then, the relation (A.5) becomes,
$$0=(m+1/s_2-s_2) (-E_i E_{i+1} E_i+E_{i+1} E_i E_{i+1}).\e$$
Now, since $m=-2 i\sin(\zeta_1)=s_2-1/s_2$, this equation is seen to hold.
The relation (A.14) is very similar and is shown to hold in the same way.

Consider now the relation eq. (A.7). Again we substitute $G_j^{-1}\rarrow -m+m E_j+G_j$ for
$j=i$ or $i+1$. We find using the relation $E_i E_{i\pm1} E_i=E_i$ (the Temperley Lieb relation)
that this equation is zero, if and only if,
$$E_i G_{i\pm1} E_i=l E_i.\e$$

This proves one direction of our assertion. Namely, that if the Yang Baxter equation holds then the
BMW algebra follows. This is because the BMW algebra is generated by
precisely the relations we found \r{\BirWen,\Mur}. These are:

1) The skein relation: $G_i-G_i^{-1}=m(1-E_i)$.

2) Braid relations: $G_i G_j=G_j G_i$ if $|i-j|\geq 2$, and $G_i G_{i+1} G_i=G_{i+1} G_i G_{i+1}$.

3) Tangle relations: $E_i E_{i\pm1} E_i=E_i$ and $G_{i\pm1} G_i E_{i\pm1}=E_i G_{i\pm1} G_i$.

4) Delooping relations: $G_i E_i=E_i G_i=l^{-1} E_i$ and $E_i G_{i\pm1} E_i=l E_i$.

Precisely, these relations we found to hold and thus we proved that the Birman--Murakami--Wenzl 
algebra holds if the Yang Baxter equation holds.

The rest of the relations can be seen (with some effort) to hold if the BMW algebra holds.
The calculations are rather involved and we omit them. This proves that the YBE holds 
if the BMW algebra is obeyed. We arrive at the conclusion that the Yang Baxter equation is fulfilled,
if and only if, the Birman--Murakami--Wenzl algebra is obeyed. Thus, we proved that any three block 
integrable lattice
model obeys the BMW algebra and it is integrable if the BMW algebra holds.

\mysec{Four blocks lattice IRF models}
We turn now to the four block case. The algebra that governs this models is termed 
4--CB (4 Conformal Braiding) algebra. We do not know all the relations of the 4--CB algebra yet. 
However, the relations that we know are enough to span the algebra.

We denote by BMW$^\prime$ the BMW as described earlier, eq. (3.1), 
along with the Temperley Lieb algebra, eq. (3.2) (with a different coefficient $b$),
with all the relations,
{\rm except}, the skein relation. Our first claim is that this algebra  BMW$^\prime$, is a sub--algebra
of the 4--CB algebra. In other words, the BMW algebra is obeyed, except, obviously, the skein 
relation which is different. 
The BMW$^\prime$ algebra is generated by $G_i$, $G_i^{-1}$ and $E_i$, which are defined below.
In fact, we conjecture that the BMW$^\prime$ algebra is obeyed 
by any number of blocks greater than two, or, it is a sub--algebra of the $n$--CB algebra 
for $n\geq 3$, again generated by $E_i$, $G_i$ and $G_i^{-1}$, which are defined similarly.
This in analogy to the Temperley--Lieb algebra which, as we proved, eq. (3.9,3.10), holds for any theory
with $n\geq2$ blocks, generated by $E_i$.

As before, we define the 4--CB algebra as the algebra generated by $G_i$ and $E_i$ where 
$$\manyeq{G_i&=8\left[\prod_{r=0}^ 2 \sin(\zeta_r)\right ] e^{-3 i\lambda/2} X_i,\cr
                   G_i^{-1}&=8\left[\prod_{r=0}^ 2 \sin(\zeta_r)\right ]  e^{3i\lambda/2} X_i^t,\cr}$$
where $\zeta_r$ were defined in eq. (2.9), $\lambda=\zeta_0$ is the crossing multiplier, 
and
$$X_i=\lim_{u\rarrow i\infty} e^{3 i u} X_i (u),\qquad X_i^t=\lim_{u\rarrow-i\infty} e^{-3 i u} X_i(u).\e$$
Again
$$E_i=X_i(\lambda).\e$$
The factor in equation (4.1) is demanded by the inversion relation, eq. (2.13), used to ensure that $G_i G_i^{-1}=1_i$.
The phase in the definition of $G_i$ is arbitrary, and is set to ensure the relations of the
BMW$^\prime$ algebra, as is seen below.

Let us consider the relations of the 4--CB algebra that we already know.
First, we have the Temperley--Lieb algebra for $E_i$ which is proved to be obeyed, eq. (3.9),
$$E_i E_{i\pm1} E_i=E_i.\e$$
In addition, we have,
$$E_i^2=b  E_i,\qquad {\rm where\ \ }  b=\prod_{r=0}^2 {\sin(\lambda+\zeta_r)\over \sin(\zeta_r)}.\e$$
This we see by substituting $u=\lambda$ in eq. (2.10) and using $(P_i^0)^2=P_i^0$.

The next relations are 
$$G_i E_i=E_i G_i=l^{-1} E_i.\e$$
These relations are verified by substituting $G_i$ from the definition, eq. (4.1), and using
$P_i^a P_i^0=P_i^0 P_i^a=\delta_{a,0} P_i^0$. We find for $l$ the value,
$$l=i e^{i(3\lambda/2+\zeta_0+\zeta_1+\zeta_2)}.\e$$ 

Now, we know the braiding relations for $G_i$:
$$G_i G_j=G_j G_i {\ \rm if\ } |i-j|\geq2, \qquad G_i G_{i+1} G_i=G_{i+1} G_i G_{i+1}.\e$$

The obvious relation not in BMW$^\prime$ is the skein relation which is,
$$G_i^2=\alpha+\beta E_i+\gamma G_i +\delta G_i^{-1},\e$$
where $\alpha,\beta,\gamma,\delta$ are constants, which depend on $\zeta_r$, $r=0,1,2$.
These constants are given in appendix (C), eq. (C.8).

From the skein relation, eq. (4.9) we find the relation
$$G_{i\pm1} G_i E_{i\pm1}=E_i G_{i\pm 1} G_i,\e$$
which follows from the BMW$^\prime$ algebra along with the skein relation.

We also find the relation,
$$E_i G_{i\pm1}^2 E_i=\kappa E_i,\e$$
which again follows from the BMW$^\prime$ algebra along with the skein relation, by substituting
the value of $G_i^2$. The coefficient 
of $\kappa$, which depends on $\zeta_i$ is determined from this and is
$$\kappa=\alpha b+\beta+\gamma l+\delta l^{-1}.\e$$
For the $SU(2)$ fused 
model (section (5)) we have $\kappa=1$, but this is not true for the general four block model.

In fact, the skein relation allows us to express $G_i^2$ in terms of $G_i$, $E_i$ and $G_i^{-1}$.
Since, as was argued above, the latter satisfy the BMW$^\prime$ algebra, these relations
are exactly enough to span the entire 4-CB algebra, i.e., the four block algebra. The additional
relations, involving $G_i^2$, which are only partially known, will be important as relations of the five block algebra.

Another consequence is that we can define a knot polynomial relation from the 4-CB algebra.
The knot polynomials are defined from words in the braid group enveloping algebra, representing
the particular knot,
see e.g. \r\GepKnot. Thus, this relations are exactly enough to reduce every such word
down to unity. This defines a new knot invariant. The details are given in appendix (D).

\mysec{The fused $3\times 3$ $SU(2)$ model}
Let us give now a concrete example of a four block IRF lattice model. This is the model
IRF$(SU(2)_k,[3],[3])$. Namely, the conformal field theory $\cal O$ is $SU(2)_k$ and the
fields $h=v=[3]$, i.e., the isospin $3/2$ representation. We denote by $l$ the isospin of the
representation and $l=0,1,2,\ldots,k$. 

The fields appearing in the fusion product of $h\times v$ are
$$[3]\times [3]=[0]+[2]+[4]+[6].\e$$
So, this is a four block theory.

The dimension formula for $SU(2)_k$ for the representation $[l]$ is
$$\Delta_l={l(l+2)\over 4(k+2)}.\e$$
The parameters $\zeta_i$ are given by
$$\zeta_0=\lambda={\pi\over 2}(\Delta_2-\Delta_0)={\pi\over k+2},\e$$
$$\zeta_1={\pi\over 2} (\Delta_4-\Delta_2)=2 \lambda,\e$$
and
$$\zeta_2={\pi\over 2}(\Delta_6-\Delta_4)=3\lambda.\e$$

The Boltzmann weights of this model are listed in appendix (B).
We checked that all the relations described by eq. (3.1,3.2), except for the skein relation, are obeyed, numerically. In particular,
we verified the BMW$^\prime$ algebra described there. The parameters $l$ and $b$ are
seen to be,
$$l=i e^{i(3\lambda/2+\zeta_0+\zeta_1+\zeta_2)}=ie^{i(15\lambda/2)},\qquad
b={\sin(4\lambda)\over \sin(\lambda)},\e$$
in accordance with eqs. (4.5,4.7).
We checked the algebra at levels $k=8,10,11$. We find a complete agreement with the 
BMW$^\prime$ algebra, as described in section (4). We were not able to check this algebra
for general $k$ due to the complexity of the calculation.

We also checked that the algebraic relations coming from expanding the YBE are all obeyed 
for this four block model. The details are given in appendix (C).

\mysec{Discussion}
In this paper, and the previous one \r\BelGep, we investigated  the 
algebraic structure of solvable lattice models. The related algebras were termed $n$-CB algebras,
where $n$ is the number of blocks. We found that the 3-CB algebra is the Birman--Murakami--Wenzl algebra for any three block theory. For the 4--CB algebra, we argued that it is 
generated by the BMW algebra with a different skein relation. Clearly, more study is needed.
In particular, deciphering the general $n$--CB algebra, for any $n$, is a major 
challenge, left to future work. Our present results indicate that the BMW algebra is
a sub--algebra of the $n$--CB algebra for any $n$, with different skein relations.

In physics, the knowledge
of the algebraic structure of solvable lattice models would contribute to the study of solvable lattice models, conformal field theory
and integrable soliton systems.

In mathematics, the studies of such models is important to knot theory, yielding new knot invariants.
In particular, as we indicated the 4-CB algebra gives a new knot invariant which was described
in appendix (D).
The IRF lattice models are also important in combinatorics, yielding new Rogers--Ramanujan type 
identities. For examples, see \r\Baxter.

 
\ack
We are extremely indebted to Ida Deichaite for her encouragement and impetus,
without which this paper would probably not have been written.  We are grateful for discussions
with B. Le Floch, H. Wenzl and J.B. Zuber.

\def\frac#1#2{{#1\over #2}}
\APPENDIX{A}{A}
\line{\bf The relations of the three block YBE algebra.\hfill}
\def\to{=}
\def\text#1{#1}
\overfullrule=0pt
We list the $19$ relations obtained by expanding the YBE in powers of $e^{iu}$ and $e^{iv}$.
We denote by $a_{i,j,k}[r,s,t]$ the element of the algebra $a_i[r] a_j[s] a_k[t]$,
where $a_i[r]$ is $G_r, G_r^{-1}$ or $E_r$ according to whether $i=1,2,3$, respectively. We define here,
$$s_1=e^{-i\zeta_0},\qquad s_2=e^{-i\zeta_1}.$$

The $19$ relations of the three block model are then calculated to be given as follows,

$$a_{2,1,1}(i,i+1,i)\to a_{1,1,2}(i+1,i,i+1),\e$$
$$a_{2,2,1}(i,i+1,i)\to a_{1,2,2}(i+1,i,i+1),\e$$
$$a_{2,2,2}(i,i+1,i)\to a_{2,2,2}(i+1,i,i+1),\e$$
$$a_{2,2,3}(i,i+1,i)\to a_{3,2,2}(i+1,i,i+1),\e$$
$$\threeline{ a_{3,3,1}(i,i+1,i)\to \left(\frac{1}{s_2}-s_2\right) \
a_{3,3,3}(i,i+1,i)+}{\left(s_2-\frac{1}{s_2}\right) \
a_{3,3,3}(i+1,i,i+1)+a_{1,3,3}(i+1,i,i+1)-a_{2,3,3}(i+1,i,i+1)+}{a_{3,3,2}(i,i\
+1,i),}$$
$$a_{3,2,2}(i,i+1,i)\to a_{2,2,3}(i+1,i,i+1),\e$$
$$\twoline{a_{3,1,3}(i,i+1,i)\to \
\frac{a_{3,2,3}(i,i+1,i)}{s_1^4}-\frac{a_{3,2,3}(i+1,i,i+1)}{s_1\
^4}+}{\frac{\left(s_2^2-1\right) a_{3,3,3}(i,i+1,i)}{s_1^2 \
s_2}+\frac{\left(1-s_2^2\right) \
a_{3,3,3}(i+1,i,i+1)}{s_1^2 s_2}+a_{3,1,3}(i+1,i,i+1),}$$
$$a_{1,2,2}(i,i+1,i)\to a_{2,2,1}(i+1,i,i+1),\e$$
$$a_{1,1,2}(i,i+1,i)\to a_{2,1,1}(i+1,i,i+1),\e$$
$$a_{1,1,1}(i,i+1,i)\to a_{1,1,1}(i+1,i,i+1),\e$$
$$a_{1,1,3}(i,i+1,i)\to a_{3,1,1}(i+1,i,i+1),\e$$
$$a_{2,3,1}(i,i+1,i)\to a_{1,3,2}(i+1,i,i+1),\e$$
$$\fiveline{a_{2,1,3}(i,i+1,i)\to -\frac{\left(s_1^2-1\right) \
\left(s_2^2-1\right)^2 a_{3,3,3}(i,i+1,i)}{s_1^2 \
s_2^2}+}{\frac{\left(s_1^2-1\right) \left(s_2^2-1\right)^2 \
a_{3,3,3}(i+1,i,i+1)}{s_1^2 \
s_2^2}-\frac{\left(s_1^2-1\right) \left(s_2^2-1\right) \
a_{1,3,3}(i,i+1,i)}{s_1^2 s_2}+}{\frac{\left(s_1^2-1\right) \
\left(s_2^2-1\right) a_{3,3,1}(i+1,i,i+1)}{s_1^2 \
s_2}-\frac{a_{1,2,3}(i,i+1,i)}{s_1^2}+\frac{a_{3,2,1}(i+1,i,i+1)\
}{s_1^2}-}{\frac{\left(s_1^2-1\right) \left(s_2^2-1\right) \
a_{3,2,3}(i,i+1,i)}{s_1^4 s_2}+\frac{\left(s_1^2-1\right) \
\left(s_2^2-1\right) a_{3,2,3}(i+1,i,i+1)}{s_1^4 \
s_2}+}{a_{3,1,2}(i+1,i,i+1),}$$
$$\twoline{a_{2,3,3}(i,i+1,i)\to \left(s_2-\frac{1}{s_2}\right) \
a_{3,3,3}(i,i+1,i)+}{\left(\frac{1}{s_2}-s_2\right) \
a_{3,3,3}(i+1,i,i+1)+a_{1,3,3}(i,i+1,i)-a_{3,3,1}(i+1,i,i+1)+a_{3,3,2}(i+1,i\
,i+1),}$$
$$\fourline{a_{3,1,2}(i,i+1,i)\to \frac{s_2 \
a_{1,2,1}(i,i+1,i)}{\left(s_1^2-1\right) \
\left(s_2^2-1\right)}-\frac{s_2 \
a_{1,2,1}(i+1,i,i+1)}{\left(s_1^2-1\right) \
\left(s_2^2-1\right)}-}{\frac{s_2 \
a_{2,1,2}(i,i+1,i)}{\left(s_1^2-1\right) \
\left(s_2^2-1\right)}+\frac{s_2 \
a_{2,1,2}(i+1,i,i+1)}{\left(s_1^2-1\right) \
\left(s_2^2-1\right)}+\frac{\left(s_1^2-1\right) \
\left(s_2^2-1\right) a_{2,3,3}(i+1,i,i+1)}{s_1^2 \
s_2}-}{\frac{\left(s_1^2-1\right) \left(s_2^2-1\right) \
a_{3,3,2}(i,i+1,i)}{s_1^2 \
s_2}+\frac{a_{1,2,3}(i,i+1,i)}{s_1^2}-\frac{a_{2,3,2}(i,i+1,i)}{\
s_1^2}+}{\frac{a_{2,3,2}(i+1,i,i+1)}{s_1^2}-\frac{a_{3,2,1}(i+1,i,\
i+1)}{s_1^2}+a_{2,1,3}(i+1,i,i+1),}$$
$$\fourline{a_{1,3,1}(i,i+1,i)\to -\frac{\left(s_1^2-1\right) \
\left(s_2^2-1\right) a_{1,3,3}(i,i+1,i)}{s_1^2 \
s_2}+}{\frac{\left(s_1^2-1\right) \left(s_2^2-1\right) \
a_{2,3,3}(i+1,i,i+1)}{s_1^2 \
s_2}+\frac{\left(s_1^2-1\right) \left(s_2^2-1\right) \
a_{3,3,1}(i+1,i,i+1)}{s_1^2 \
s_2}-}{\frac{\left(s_1^2-1\right) \left(s_2^2-1\right) \
a_{3,3,2}(i,i+1,i)}{s_1^2 \
s_2}-\frac{a_{2,3,2}(i,i+1,i)}{s_1^2}+\frac{a_{2,3,2}(i+1,i,i+1)\
}{s_1^2}+}{a_{1,3,1}(i+1,i,i+1),}$$
$$a_{1,3,2}(i,i+1,i)\to a_{2,3,1}(i+1,i,i+1),\e$$
$$\fourline{a_{3,2,1}(i,i+1,i)\to -\frac{s_1^2 s_2 \
a_{1,2,1}(i,i+1,i)}{\left(s_1^2-1\right) \
\left(s_2^2-1\right)}+\frac{s_1^2 s_2 \
a_{1,2,1}(i+1,i,i+1)}{\left(s_1^2-1\right) \
\left(s_2^2-1\right)}+}{\frac{s_1^2 s_2 \
a_{2,1,2}(i,i+1,i)}{\left(s_1^2-1\right) \
\left(s_2^2-1\right)}-\frac{s_1^2 s_2 \
a_{2,1,2}(i+1,i,i+1)}{\left(s_1^2-1\right) \
\left(s_2^2-1\right)}-\frac{\left(s_1^2-1\right) \
\left(s_2^2-1\right) a_{3,2,3}(i,i+1,i)}{s_1^2 \
s_2}+}{\frac{\left(s_1^2-1\right) \left(s_2^2-1\right) \
a_{3,2,3}(i+1,i,i+1)}{s_1^2 \
s_2}-}{a_{1,2,3}(i,i+1,i)+a_{1,2,3}(i+1,i,i+1)+a_{2,3,2}(i,i+1,i)-a_{2,3
,2}(i+1,i,i+1)+a_{3,2,1}(i+1,i,i+1),}$$
$$a_{3,1,1}(i,i+1,i)\to a_{1,1,3}(i+1,i,i+1).\e$$

\endpage

\APPENDIX{B}{B}
\line{\bf Weights of the $3\times 3$ fused model.\hfill}

These are the weights of the model IRF$(SU(2)_k,[3],[3])$. We use the notation,
$$\lambda={\pi\over k+2}\e$$
and 
$$s[x]={\sin(x)\over \sin(\lambda)}.\e$$
The weights are taken from ref. \REF\Pearce{E. Tartaglia and P. Pearce, J. Phys. A 49 (2016) 18.}\r\Pearce, based on the calculations
of ref.  \r\Jimbo. We shifted the fields $a\rarrow a+1$ so the weights 
range over $a=1,2,\ldots,k+1$, namely, $a$ is the dimension of the $SU(2)$ representation.

\def\Wthree#1#2#3#4#5{{\omega\left( \matrix{#4 & #1\cr #3 & #2 \cr}\bigg | #5  \right)}}
$$\manyeq{
\Wthree{a}{a\pm3}{a}{a\pm3}{u}&=\frac{s((a\pm1)\lambda\mp u) s((a\pm2) \lambda \mp u) s((a\pm3) \lambda \mp u)}{s((a\pm1) \lambda )s((a\pm2) \lambda ) s((a\pm3) \lambda )}\cr
\Wthree{a}{a\pm3}{a}{a\mp3}{u}&=\frac{s(\lambda -u) s(2 \lambda -u) s(3 \lambda-u)}{s(2 \lambda ) s(3 \lambda )}\cr
\Wthree{a}{a\pm3}{a}{a\pm1}{u}&=\Wthree{a}{a\pm1}{a}{a\pm3}{u}=\frac{s(\lambda -u)   s((a\pm1) \lambda\mp u) s((a\pm2) \lambda \mp u)}{s((a\pm 1) \lambda ) s((a\pm2) \lambda)}\cr
\Wthree{a}{a\pm3}{a}{a\mp1}{u}&=\Wthree{a}{a\mp1}{a}{a\pm3}{u}=\frac{s(\lambda -u) s(2 \lambda -u) s((a\pm1) \lambda\mp u)}{s(2 \lambda ) s((a\pm1) \lambda )}\cr
\Wthree{a}{a\pm1}{a}{a\pm1}{u}&=\frac{s((a\pm1) \lambda\mp u) s((a\pm 1) \lambda\pm u) s((a\pm2) \lambda \mp u)}{s((a\pm1) \lambda )^2 s((a\pm2) \lambda )}\cr
&-\frac{s(2 \lambda) s((a-2) \lambda ) s((a+2) \lambda )s(\lambda - u)  s(u) s((a\pm1) \lambda\mp u)}{s(3 \lambda ) s((a\mp1) \lambda ) s((a\pm1) \lambda   )^2}\cr
      \Wthree{a}{a\pm1}{a}{a\mp1}{u}&=\frac{s(2 \lambda )^2 s((a\mp2) \lambda ) s(\lambda - u) s(a \lambda \pm u) s((a\pm1) \lambda \mp u)}{s(3 \lambda ) s((a\mp1) \lambda )^2   s((a\pm1) \lambda )}\cr
        &-\frac{s((a\mp3) \lambda ) s((a\pm1) \lambda ) s(2 \lambda -u) s(\lambda - u) s(\lambda +u) }{s(2 \lambda ) s(3 \lambda   ) s((a\mp1) \lambda )^2}\cr
}$$
$$\eqalign{
\Wthree{a}{a\pm3}{a\pm6}{a\pm3}{u}&=-\frac{s((a\pm4) \lambda ) s((a\pm5) \lambda ) s((a\pm6) \lambda )s(u)  s(\lambda +u) s(2 \lambda +u)}{s(2 \lambda ) s(3 \lambda ) s((a\pm1) \lambda ) s((a\pm2) \lambda ) s((a\pm3) \lambda )}\cr
\Wthree{a}{a\pm3}{a\pm4}{a\pm3}{u}&=\frac{s((a\pm4) \lambda ) s((a\pm5) \lambda ) s(u) s(\lambda +u) s((a\pm3) \lambda \mp u)}{s(2 \lambda ) s(3 \lambda ) s((a\pm1) \lambda ) s((a\pm2) \lambda ) s((a\pm3) \lambda )}\cr
\Wthree{a}{a\pm3}{a\pm4}{a\pm1}{u}&=\Wthree{a}{a\pm1}{a\pm4}{a\pm3}{u}=-\frac{s((a\pm4) \lambda ) s((a\pm5) \lambda ) s(u) s(u-\lambda ) s(\lambda +u)}{s(2 \lambda ) s(3 \lambda ) s((a\pm1) \lambda ) s((a\pm2) \lambda )}\cr
\Wthree{a}{a\pm3}{a\pm2}{a\pm3}{u}&=-\frac{s((a\pm4) \lambda ) s(u) s((a\pm2) \lambda \mp u) s((a\pm3) \lambda \mp u)}{s(3 \lambda ) s((a\pm 1) \lambda ) s((a\pm2) \lambda ) s((a\pm3) \lambda )}\cr
\Wthree{a}{a\pm3}{a\pm2}{a\pm1}{u}&=\Wthree{a}{a\pm1}{a\pm2}{a\pm3}{u}=\frac{s((a\pm4) \lambda ) s(u) s(u-\lambda ) s((a\pm2) \lambda \mp u)}{s(3 \lambda ) s((a\pm1) \lambda ) s((a\pm2) \lambda )}\cr
\Wthree{a}{a\pm3}{a\pm2}{a\mp1}{u}&=\Wthree{a}{a\mp1}{a\pm2}{a\pm3}{u}=-\frac{s((a\pm4) \lambda ) s(2 \lambda-u ) s(\lambda-u )s(u)}{s(2 \lambda ) s(3 \lambda ) s((a\pm1) \lambda )}\cr
}$$
$$\eqalign{
\Wthree{a}{a\pm1}{a\pm4}{a\pm1}{u}&=\frac{s(3\lambda)s((a\pm3)\lambda)(s(a\pm4)\lambda)s(u)s(u+\lambda)s((a\pm 1)\lambda\pm u)}{s(2\lambda)s((a-1)\lambda)s((a+1)\lambda)s((a\pm2)\lambda)}\cr
\Wthree{a}{a\pm1}{a\pm2}{a\pm1}{u}&=-\frac{s(a \lambda ) s((a\pm3) \lambda ) s((a\pm4) \lambda ) s(u)^2 s(u-\lambda )}{s(2 \lambda ) s(3 \lambda ) s((a\pm1) \lambda )^2 s((a\pm2) \lambda )}\cr
&-\frac{s((a\pm3) \lambda )s(u)  s(a \lambda \pm u) s((a\pm1) \lambda\mp u)}{s((a\mp1) \lambda ) s((a\pm1) \lambda )^2}\cr
\Wthree{a}{a\pm1}{a\pm2}{a\mp1}{u}&=\Wthree{a}{a\mp1}{a\pm2}{a\pm1}{u}=\frac{s((a\pm3)\lambda)s(u)s(u-\lambda)s(a\lambda\pm u)}{s((a-1)\lambda)s((a+1)\lambda)}\cr
\Wthree{a}{a\pm1}{a\mp2}{a\pm1}{u}&=-\frac{s(3\lambda)s((a\mp 2)\lambda)s(u)s(a\lambda\mp u)s((a\pm 1)\lambda\mp u)}{s((a-1)\lambda)s((a+1)\lambda)s((a\pm 2)\lambda)}
\cr}$$
\endpage

\APPENDIX{C}{C}
\line{\bf The algebraic expansion of the four block YBE.\hfill}

According to the conjecture in \r\Found~(checked recently for 3-block case in \r\BelGep ), the trigonometric solution of the YBE, eq. (2.3)
is given by eq. (2.10). The generators of the desired algebra are defined to be proportional to the limiting values of $X_i$ and $X_i^t$, arising in the limit $u\rightarrow \pm i\infty$. 

Hence, in the four-block case we can identify 
$$
G_i=8 e^{\frac{1}{2} (-(3 i)) \zeta_0} \sin (\zeta_0) \sin (\zeta_1) \sin (\zeta_2) \left(\lim_{\text{u}\rarrow  i \infty } \, \exp (3 i \text{u}) X_i(\text{u})\right) \e
$$
$$
G_i^{-1}=8 e^{\frac{1}{2} (3 i) \zeta_0} \sin (\zeta_0) \sin (\zeta_1) \sin (\zeta_2) \left(\lim_{\text{u}\rarrow -i \infty } \, \exp (-3 i \text{u}) X_i(\text{u})\right) \e
$$
$$
E_i=X_i(\zeta_0) \e
$$
Taking into account the properties of the projectors,  eq. (2.8),
 we can also introduce forth relation for the generator $G^2_i$.
 
 Explicitly we have
 $$
\twoline{
G_i=i e^{-\frac{5}{2} i \zeta_0-i \zeta_1-i \zeta_2} \big(e^{2 i \zeta_0} P_i^1-e^{2 i \zeta_0+2 i \zeta_1} P_i^2+}{e^{2 i \zeta_0+2 i \zeta_1
+2 i \zeta_2} P_i^3
-P_i^0\big),}
 $$
 $$
\twoline{
G^{-1}_i=i e^{\frac{1}{2} i \zeta_0-i \zeta_1-i \zeta_2} \big(e^{2 i \zeta_0
+2 i \zeta_1+2 i \zeta_2} P_i^0-}{e^{2 i \zeta_1+2 i \zeta_2} \
P_i^1+e^{2 i \zeta_2} P_i^2-P_i^3\big),}
$$
 $$
 \twoline{ E_i=
\frac{e^{-3 i \zeta_0} \left(1+e^{2 i \zeta_0}\right) \left(-1+e^{i \zeta_0+i \zeta_1}\right) \left(1+e^{i \zeta_0+i \zeta_1}\right)}{\left(-1+e^{i \zeta_1}\right) \left(1+e^{i \zeta_1}\right) \left(-1+e^{i \zeta_2}\right) \left(1+e^{i \zeta_2}\right)} \times}{
\left(-1+e^{i \zeta_0+i \zeta_2}\right)  \left(1+e^{i \zeta_0+i \zeta_2}\right) P_i^0,}
 $$
$$
\twoline{G_i^2=
-e^{-5 i \zeta_0-2 i \zeta_1-2 i \zeta_2} P_i^0-e^{-i \zeta_0-2 i \
\zeta_1-2 i \zeta_2} P_i^1-}{e^{-i \zeta_0+2 i \zeta_1-2 i \zeta_2} \
P_i^2-e^{-i \zeta_0+2 i \zeta_1+2 i \zeta_2} P_i^3.}
$$

That is we have the system of four linear equations, which allows to express four projectors in terms of four generators $G_i$, $G^{-1}_i$, $E_i$ and $G^2_i$. Using these expressions for projectors and eq. (2.10),  we get $X_i(u)$ expressed in terms of the desired algebra generators.

Finally, we note that the dependence on the spectral parameters $u$, $v$  in the YBE equation, enters only through the coefficients $f_a(u)$. So that The YBE  becomes a polynomial equation in the two variables $e^{i u}$ and $e^{i v}$, which is equivalent to the requirement that
all the coefficients are equal to zero. This gives a set of three-linear relations for the new algebra generators.

By using the equation $\sum_a P^a_i=1$ we get the skein relation expressing $G_i^2$ 
in terms of $G_i$, $E_i$ and $G_i^{-1}$. The skein equation is then seen to be,
$$\threeline{
G_i^2= i e^{-\frac{1}{2} i \zeta_0-i \zeta_1-i \zeta_2} \left(1-e^{2 i \zeta_1}+e^{2 i \zeta_1+2 i \zeta_2}\right) \
G_i+i e^{-\frac{3}{2} i \zeta_0+i \zeta_1-i \zeta_2} \
G_i^{-1}}{+\frac{e^{-2 i \zeta_0-2 i \zeta_1-2 i \zeta_2} \left(e^{2 i 
\zeta_1}-1\right) \left(1+e^{2 i \zeta_0+2 
i \zeta_1+2 i \zeta_2}\right) \left(e^{2 i \zeta_2}-1\right) 
 }{\left(e^{2 i \zeta_0+ 2 i \zeta_2}-1\right) }E_i}{-e^{-i \zeta_0-2 i 
\zeta_2} \left(1-e^{2 i \zeta_2}+e^{2 i \zeta_1+2 i \zeta_2}\right).}$$
We define the coefficients $\alpha$, $\beta$, $\gamma$ and $\delta$ by equating eq. (C.8)
to
$$G_i^2=\alpha+\beta E_i+\gamma  G_i+\delta G_i^{-1}.\e$$

\endpage

\APPENDIX{D}{D}
\line{\bf New knot invariants.\hfill}

We define an invariant on a link diagram $K$ as follows,
$$
\upsilon  (K) =l^{w(K)} L(K),
$$
where $w(K)$ is the writhe of the link $K$ which is defined as the number of left crossings minus the number of right crossings, and $l$ is given by eq.(4.7) and is a parameter.

We define the link function $L(K)$ as follows,

1) $L(0)=1,$

2)  $L(S_r)= l^{-1}  L(S)$ and  $L(S_l) = l L(S),$

3)  $L$ is unchanged under type II, III Reidemeister moves, see fig.(2).


\noindent Here $0$ is the unknot, $S$ is a strand and $S_r$ (respectively $S_l$) is the same strand with a right-handed (respectively left-handed) curl added, as in type I Reidemeister move.
\vskip 1cm
\center
\tikzpicture[scale=1]
\draw [line width=1pt] (32,0) -- (32,1.2);
\draw [line width=1pt] (32,0) -- (33.5,0);
\draw [line width=1pt] (33.5,0) -- (33.5,.8);
\draw [line width=1pt] (32.8,.8) -- (33.5,.8);
\draw [line width=1pt] (32.8,.8) -- (32.8,.1);
\draw [line width=1pt] (32.8,-.1) -- (32.8,-.5);
\draw [line width=1pt] (34.8,-.5) -- (34.8,1.2);
\draw (34.2,.3) node {$\rightarrow$};
\draw (33.5,1.9) node {Type I};
\draw [line width=1pt] (37,0) -- (37,-.5);
\draw [line width=1pt] (37,1.2) -- (37,.8);
\draw [line width=1pt] (37,.8) -- (37.7,.8);
\draw [line width=1pt] (37,0) -- (37.7,0);
\draw [line width=1pt] (37.9,.8) -- (38.5,.8);
\draw [line width=1pt] (37.9,0) -- (38.5,0);
\draw [line width=1pt] (38.5,0) -- (38.5,.8);
\draw [line width=1pt] (37.8,-.5) -- (37.8,1.2);
\draw [line width=1pt] (39.8,-.5) -- (39.8,1.2);
\draw [line width=1pt] (40.4,-.5) -- (40.4,1.2);
\draw (39.2,.3) node {$\rightarrow$};
\draw (38.8,1.9) node {Type II};
\draw [line width=1pt] (42,0) -- (42,-.5);
\draw [line width=1pt] (42,0) -- (43.,0);
\draw [line width=1pt] (43.,0) -- (43.,1.2);
\draw [line width=1pt] (42.,.8) -- (42.9,.8);
\draw [line width=1pt] (43.1,.8) -- (43.5,.8);
\draw [line width=1pt] (42.5,.1) -- (42.5,.7);
\draw [line width=1pt] (42.5,.9) -- (42.5,1.2);
\draw [line width=1pt] (42.5,-.1) -- (42.5,-.5);
\draw [line width=1pt] (44.8,0) -- (45.2,0);
\draw [line width=1pt] (45.4,0) -- (46.3,0);
\draw [line width=1pt] (45.3,-.5) -- (45.3,.8);
\draw [line width=1pt] (45.3,.8) -- (46.3,.8);
\draw [line width=1pt] (46.3,.8) -- (46.3,1.2);
\draw [line width=1pt] (45.8,.1) -- (45.8,.7);
\draw [line width=1pt] (45.8,.9) -- (45.8,1.2);
\draw [line width=1pt] (45.8,-.1) -- (45.8,-.5);
\draw (44.2,.3) node {$\rightarrow$};
\draw (44.3,1.9) node {Type III};
\endtikzpicture
\endcenter
\vskip 5mm
\center
\Fig 1. {Reidemeister moves.}
\endcenter
\endpage

In addition $L$ obeys the skein relations,
\vskip 1cm
\center
\tikzpicture[scale=1]
\draw [rounded corners,dashed, line width=.8pt](-.9,-1.05) rectangle (.9,1.05) (0,0);
\draw [line width=1pt](0,0) ellipse (.32 and .22);
\draw  [line width=1pt] (-.6,1.) to[out=-80,in=180] node [sloped,above] {} (0,.5);
\draw  [line width=1pt] (0,.5) to[out=0,in=-100] node [sloped,above] {} (.6,1.);
\draw  [line width=1pt] (-.6,-1.) to[out=80,in=180] node [sloped,above] {} (0,-.5);
\draw  [line width=1pt] (0,-.5) to[out=0,in=100] node [sloped,above] {} (.6,-1.);
\draw (0,1.5) node {$E^2_i$};
\draw (1.35,0) node {$=$};
\draw (1.8,0) node {$b$};
\draw (3,1.5) node {$E_i$};
\draw [rounded corners,dashed, line width=.8pt](3.-.9,-1.05) rectangle (3.+.9,1.05) (0,0);
\draw  [line width=1pt] (3.-.6,1.) to[out=-80,in=180] node [sloped,above] {} (3.+0,.3);
\draw  [line width=1pt] (3.+0,.3) to[out=0,in=-100] node [sloped,above] {} (3.+.6,1.);
\draw  [line width=1pt] (3.+-.6,-1.) to[out=80,in=180] node [sloped,above] {} (3.+0,-.3);
\draw  [line width=1pt] (3.+0,-.3) to[out=0,in=100] node [sloped,above] {} (3.+.6,-1.);
\endtikzpicture
\endcenter

\noindent and 
\vskip 1cm
\center
\tikzpicture[scale=1]
\draw [rounded corners,dashed, line width=.8pt](-.9,-1.05) rectangle (.9,1.05) (0,0);
\draw [line width=1pt] (-.6,-1.) -- (0.32,0.);
\draw [line width=1pt] (.6,1.) -- (-0.32,0.);
\draw [line width=1pt] (-.6,1.) -- (-0.1,0.45);
\draw [line width=1pt] (.6,-1.) -- (0.1,-0.45);
\draw [line width=1pt] (-0.1,-0.24) -- (-0.32,0.);
\draw [line width=1pt] (0.1,0.24) -- (0.32,0.);
\draw (0,1.5) node {$G_i^2$};
\draw (1.35,0) node {$=$};
\draw (1.8,0) node {$\alpha$};
\draw (3,1.5) node {{\bf 1}$_i$};
\draw [rounded corners,dashed, line width=.8pt](3.-.9,-1.05) rectangle (3.+.9,1.05) (0,0);
\draw  [line width=1pt] (3.-.7,.9) to[out=-30,in=30] node [sloped,above] {} (3.-.7,.-.9);
\draw  [line width=1pt] (3.+.7,.9) to[out=-150,in=150] node [sloped,above] {} (3.+.7,.-.9);
\draw (4.35,0) node {$+$};
\draw (4.8,0) node {$\beta$};
\draw (6,1.5) node {$E_i$};
\draw [rounded corners,dashed, line width=.8pt](6.-.9,-1.05) rectangle (6.+.9,1.05) (0,0);
\draw  [line width=1pt] (6.-.6,1.) to[out=-80,in=180] node [sloped,above] {} (6.+0,.3);
\draw  [line width=1pt] (6.+0,.3) to[out=0,in=-100] node [sloped,above] {} (6.+.6,1.);
\draw  [line width=1pt] (6.+-.6,-1.) to[out=80,in=180] node [sloped,above] {} (6.+0,-.3);
\draw  [line width=1pt] (6.+0,-.3) to[out=0,in=100] node [sloped,above] {} (6.+.6,-1.);
\draw (7.35,0) node {$+$};
\draw (7.8,0) node {$\gamma$};
\draw (9,1.5) node {$G_i$};
\draw [rounded corners,dashed, line width=.8pt](9.-.9,-1.05) rectangle (9.+.9,1.05) (0,0);
\draw [line width=1pt] (9.-.6,1.) -- (9.0-0.08,0.12);
\draw [line width=1pt] (9.+.6,-1.) -- (9.0+0.08,-0.12);
\draw [line width=1pt] (9.-.6,-1.) -- (9.+.6,1.);
\draw (10.35,0) node {$+$};
\draw (10.8,0) node {$\delta$};
\draw (12,1.5) node {$G^{-1}_i$};
\draw [rounded corners,dashed, line width=.8pt](12.-.9,-1.05) rectangle (12.+.9,1.05) (0,0);
\draw [line width=1pt] (12.-.6,-1.) -- (12.0-0.08,-0.12);
\draw [line width=1pt] (12.+.6,+1.) -- (12.0+0.08,+0.12);
\draw [line width=1pt] (12.-.6,+1.) -- (12.0+0.6,-1.);
\endtikzpicture
\endcenter

\noindent where $\alpha,\beta,\gamma,\delta$ are given by eq.(C.8) and $b$ by  eq. (4.5). It also
obeys, from the skein relation, eq. (C.8,C.9),
$$
b=1/\beta \left(1/l^2 - \alpha - \gamma/l - \delta l\right).
$$ 

This is a three parameter tangle algebra depending on  $\zeta_0,\zeta_1,\zeta_2$. This tangle algebra is isomorphic to the BMW$^\prime$ algebra. The isomorphism is given by,
\vskip 1cm
\center
\tikzpicture[scale=1]
\draw (34.2,.3) node {$G_i\,\,\longmapsto$};
\draw [line width=1pt] (35.4,-.5) -- (35.4,1.2);
\draw (35.8,.3) node {$\dots$};
\draw [line width=1pt] (36.2,-.5) -- (36.2,1.2);
\draw [line width=1pt] (37.,-.5) -- (37.8,1.2);
\draw [line width=1pt] (37.,1.2) -- (37.33,.5);
\draw [line width=1pt] (37.48,.2) -- (37.8,.-.5);
\draw [line width=1pt] (38.6,-.5) -- (38.6,1.2);
\draw (39.,.3) node {$\dots$};
\draw [line width=1pt] (39.4,-.5) -- (39.4,1.2);
\draw (35.4,1.5) node {1};
\draw (36.2,1.5) node {i-1};
\draw (37.,1.5) node {i};
\draw (37.8,1.5) node {i+1};
\draw (38.6,1.5) node {i+2};
\draw (39.4,1.5) node {n};
\draw (40.5,.3) node {and};
\draw (42.2,.3) node {$E_i\,\,\longmapsto$};
\draw [line width=1pt] (43.4,-.5) -- (43.4,1.2);
\draw (43.8,.3) node {$\dots$};
\draw [line width=1pt] (44.2,-.5) -- (44.2,1.2);
\draw [line width=1pt] (46.6,-.5) -- (46.6,1.2);
\draw (47.,.3) node {$\dots$};
\draw [line width=1pt] (47.4,-.5) -- (47.4,1.2);
\draw (43.4,1.5) node {1};
\draw (44.2,1.5) node {i-1};
\draw (45.,1.5) node {i};
\draw (45.8,1.5) node {i+1};
\draw (46.6,1.5) node {i+2};
\draw (47.4,1.5) node {n};
\draw  [line width=1pt] (45.,1.2) to[out=-80,in=180] node [sloped,above] {} (45.4,.65);
\draw  [line width=1pt] (45.4,.65) to[out=0,in=-100] node [sloped,above] {} (45.8,1.2);
\draw  [line width=1pt] (45.,-.5) to[out=80,in=180] node [sloped,above] {} (45.4,.1);
\draw  [line width=1pt] (45.4,.1) to[out=0,in=100] node [sloped,above] {} (45.8,-.5);
\endtikzpicture
\endcenter
\center
\endcenter

The BMW$^\prime$ algebra ensures invariance under Reidemeister moves and skein relations.

Using this tangle algebra any knot invariant can be calculated. The fact that this $L(K)$ exists and is a regular isotopy
invariant follows from the consistency of the BMW$^\prime$ algebra, for which we have explicit representation  for some $\zeta_i$,
which correspond to some solvable lattice model. 
Thus, $\upsilon(K)$ is an invariant (ambient isotopy invariant) of oriented links.
For the general values of the parameters,
we did not prove the consistency of the BMW$^\prime$ algebra, and this is left to further work.

This defines a three parameter link invariant. It is, in fact, the same invariant defined through the Boltzmann weights in ref. \r{\GepKnot,\Wadati}, which we term the IRF invariant. The advantage of our approach is the following. First, our link invariant can be calculated by the skein relations, unlike the IRF knot invariant where one cannot express $G_i^2$, for four block theories. 
Second, our invariant holds for  all the values of the parameters $\zeta_0,\zeta_1,\zeta_2$, and is thus a three parameter link invariant, whereas
the IRF invariant is special to such values of the parameters appearing in conformal field theory.

The benefit of our three parameter link invariant is that it could be used to distinguish links
which cannot be told apart by existing link invariants.

\refout

\bye

%% file: mydef.tex
\def\e{\adveq\eqno{\rm (\chapterlabel\the\equanumber)}}
\def\mysec#1{\equanumber=0\chapter{#1}}
\def\adveq{\global\advance\equanumber by 1}
\def\myeq{{\rm \chapterlabel.\the\equanumber}}
\def\rarrow{\rightarrow}

\def\twoline#1#2{\displaylines{\qquad#1\hfill(\adveq\myeq)\cr\hfill#2
\qquad\cr}}

\def\manyeq#1{\eqalign{#1}\e}

\def\semidirect{\mathrel{\raise0.04cm\hbox{${\scriptscriptstyle |\!}$
\hskip-0.175cm}\times}}


\def\ref#1{$^{[#1]}$}

\def\r#1{$[\rm#1]$}

\def\threeline#1#2#3{\displaylines{\qquad#1\hfill\cr\hfill#2\hfill\llap{(\adveq\myeq)}\cr
\hfill#3\qquad\cr}}

\def\e{\adveq\eqno{\rm (\chapterlabel.\the\equanumber)}}
\def\mysec#1{\equanumber=0\chapter{#1}}
\def\adveq{\global\advance\equanumber by 1}
\def\myeq{{\rm \chapterlabel.\the\equanumber}}
\def\rarrow{\rightarrow}

\def\twoline#1#2{\displaylines{\qquad#1\hfill(\adveq\myeq)\cr\hfill#2
\qquad\cr}}

\def\manyeq#1{\eqalign{#1}\e}

\def\semidirect{\mathrel{\raise0.04cm\hbox{${\scriptscriptstyle |\!}$
\hskip-0.175cm}\times}}


\def\ref#1{$^{[#1]}$}

\def\r#1{$[\rm#1]$}

\def\threeline#1#2#3{\displaylines{\qquad#1\hfill\cr\hfill#2\hfill\llap{(\adveq\myeq)}\cr
\hfill#3\qquad\cr}}

\def\threeline#1#2#3{\displaylines{\qquad#1\hfill\llap{(\adveq\myeq)}\cr\hfill#2\hfill \cr
\hfill#3\qquad\cr}}
\def\fourline#1#2#3#4{\displaylines{#1\hfill\llap{(\adveq\myeq)}\cr#2\hfill \cr\hfill#3\hfill\cr
\hfill#4\qquad\cr}}
\def\fiveline#1#2#3#4#5{\displaylines{#1\hfill\llap{(\adveq\myeq)}\cr#2\hfill \cr\hfill#3\hfill\cr\hfill#4\cr
\hfill#5\qquad\cr}}